\documentclass[journal,a4paper]{IEEEtran}

\IEEEoverridecommandlockouts

\usepackage[english]{babel}
\usepackage{amsmath}
\usepackage{graphicx}
\usepackage[colorlinks=true, allcolors=blue]{hyperref}
\usepackage{subfiles}
\usepackage{soul}
\usepackage{subcaption}
\usepackage{epstopdf}
\usepackage{xcolor}
\usepackage{multirow}
\usepackage{booktabs}

\def\BibTeX{{\rm B\kern-.05em{\sc i\kern-.025em b}\kern-.08em
    T\kern-.1667em\lower.7ex\hbox{E}\kern-.125emX}}

\usepackage{verbatim}

\title{Machine Learning \& Wi-Fi: Unveiling the Path Towards AI/ML-Native IEEE 802.11 Networks}
\author{Francesc Wilhelmi, Szymon Szott, Katarzyna Kosek-Szott, Boris Bellalta\thanks{F. Wilhelmi is with Nokia Bell Labs. S. Szott and K. Kosek-Szott are with AGH University of Krakow. B. Bellalta is with Universitat Pompeu Fabra.\\This paper is supported by the CHIST-ERA Wireless AI 2022 call MLDR project (ANR-23-CHR4-0005), partially funded by AEI and NCN under projects PCI2023-145958-2 and 2023/05/Y/ST7/00004, respectively. B. Bellalta's contribution is supported by Wi-XR PID2021123995NB-I00 (MCIU/AEI/FEDER,UE) and MdMCEX2021-001195-M/ AEI /10.13039/501100011033.}}

\begin{document}

\bstctlcite{IEEEexample:BSTcontrol}

\maketitle

\begin{abstract}
Artificial intelligence (AI) and machine learning (ML) are nowadays mature technologies considered essential for driving the evolution of future communications systems. Simultaneously, Wi-Fi technology has constantly evolved over the past three decades and incorporated new features generation after generation, thus gaining in complexity. As such, researchers have observed that AI/ML functionalities may be required to address the upcoming Wi-Fi challenges that will be otherwise difficult to solve with traditional approaches. This paper discusses the role of AI/ML in current and future Wi-Fi networks and depicts the ways forward. A roadmap towards AI/ML-native Wi-Fi, key challenges, standardization efforts, and major enablers are also discussed. An exemplary use case is provided to showcase the potential of AI/ML in Wi-Fi at different adoption stages.
\end{abstract}


\section{Introduction}
\label{sec:introduction}

\begin{figure*}[ht!] 
    \centering
    \includegraphics[width=0.9\textwidth]{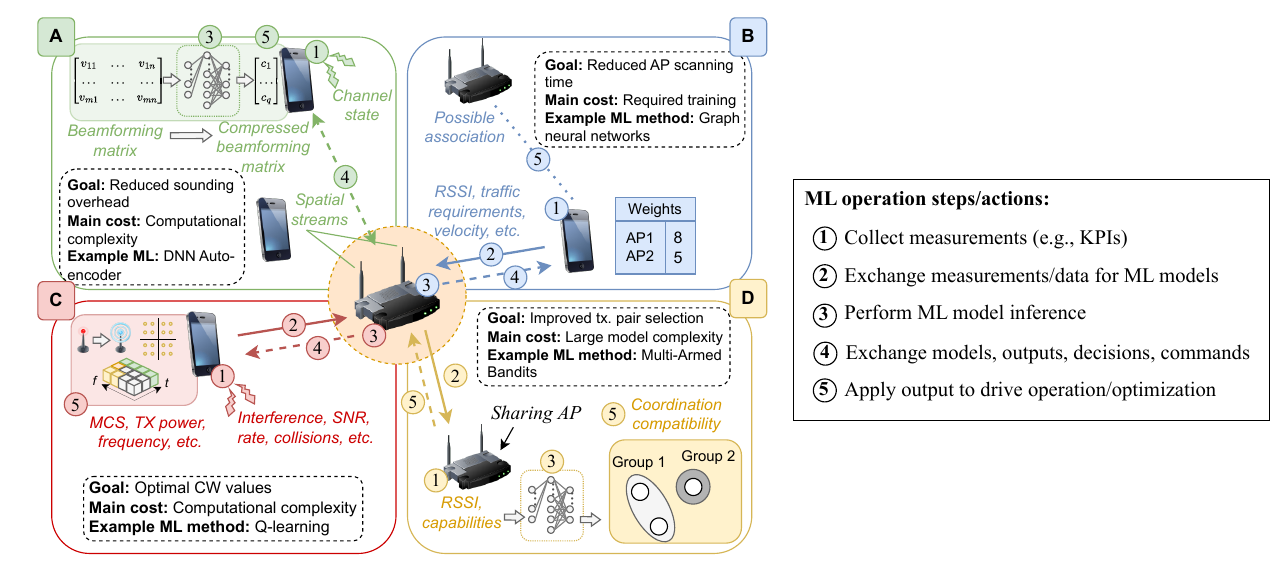}
    \caption{Overview of the AIML TIG use cases: (A) CSI feedback compression, (B) Enhanced roaming, (C) DRL-based channel access, (D) MAPC driven by AI/ML.}
    \label{fig:use_cases}
\end{figure*}

The IEEE 802.11 standard (commercially known as \mbox{Wi-Fi}), since its first release in 1997, has accompanied humanity by providing an entry point to a connected world. Wi-Fi has continuously evolved together with its users, their applications, and habits. As a result, it continues to be the most popular technology for indoor connectivity \cite{edirisinghe2023recent}\footnote{Even if cellular networks provide stationary Internet connectivity through solutions like Fixed Wireless Access (FWA), the indoor part of the access network remains Wi-Fi. Meanwhile, future cellular 5G/6G networks may also be deployed indoors, although given the proliferation of Wi-Fi, such deployments will most likely be restricted to private networks in industrial environments.}. In addition, the technological evolution of Wi-Fi---which has included a plethora of physical~(PHY) and medium access control~(MAC) enhancements---has enabled it to expand to other domains beyond residential, including enterprise and industry.

Today, communication systems are expected to enable new verticals like Industry 5.0, autonomous vehicles, or smart cities, where unprecedented levels of performance and reliability are required. However, the current incremental-based status quo on evolving wireless communications standards and technologies---with 5G/6G as two of the most relevant examples with the introduction of new paradigms like ultra-reliable low latency communications (uRLLC) or integrated sensing and communications---seems to be insufficient for such stringent performance targets. Moreover, the underlying complexity and the incurred overheads of some mechanisms adopted in communications (e.g., massive antenna arrays) demand innovative ways of operating networks, which must now hoard self-configuration capabilities~\cite{viswanathan2020communications}. 

Nowadays, we are witnessing the first steps towards the adoption of Artificial Intelligence (AI) and Machine Learning (ML) in communication networks to automate high-level management operations such as the identification and prediction of potential issues and failures. However, the aforementioned factors call for a paradigm shift, which could be achieved by AI/ML. AI/ML is a computer programming paradigm that allows learning functions directly from data. AI/ML has boosted many domains thanks to its ability to exploit complex characteristics from data, thus allowing it to solve problems that are hard to solve by hand. So now the question is to what extent can AI/ML reshape communications. Will AI/ML be applied to support the network's operation? Will it be part of current communications protocols? Or will it create new protocols?

For wireless networks, AI/ML is intended as a key building block of 5G/6G networks \cite{shehzad2022artificial}.  3GPP is incorporating AI/ML in its cellular radio standards \cite{lin2024overview}.
Meanwhile, in recent years, the utilization of AI/ML in Wi-Fi has also grown exponentially. In academic research, AI/ML has been extensively studied to showcase potential, provide a better understanding of its trade-offs, and foresee its viability when applied to Wi-Fi use cases~\cite{szott2022wifi}. In industry, AI/ML can be found in commercial Wi-Fi solutions as a product differentiator, mostly oriented to enhance the operator's management experience (e.g., network analytics and assisted troubleshooting) or to provide basic self-configuration (e.g., automatic channel selection). AI-powered solutions like HPE Aruba Networking Central are based on proprietary implementations, some of which leverage non-AI/ML-specialized standardized protocols---such as the Broadband Forum (BBF) User Services Platform (USP) data models---to enable data collection and remote device configuration. The main drawback of these kinds of proprietary solutions is that they are far from the Wi-Fi core, thus their potential for improving Wi-Fi is limited.

In view of the high expectations of AI/ML for Wi-Fi, in this paper, we shed light on the status and future evolution of the 802.11 tied to AI/ML, starting from current developments to potential AI nativeness. We identify the main challenges and enablers for pursuing such an evolutionary path, emphasizing on standardization gaps and required actions to be taken. In addition, we provide simulation results to showcase the benefits that AI/ML can in its different adoption stages bring to Wi-Fi. Based on past and future milestones on AI/ML, Fig.~\ref{fig:roadmap} depicts the envisioned roadmap for AI/ML adoption in Wi-Fi, which we later discuss in detail.

\section{Current AI/ML in IEEE 802.11 Standardization}
\label{sec:standardization}

The first standardization steps towards AI-native Wi-Fi operation were recently taken by the IEEE 802.11 AIML Topic Interest Group (TIG), which produced a technical report describing relevant use cases of AI/ML for 802.11~\cite{aiml_tig_use_cases}. This report can influence other groups and standards within 802.11, particularly 802.11bn (\mbox{Wi-Fi}~8), hence potentially leading to the definition of new mechanisms, signaling, infrastructure, and interfaces to enable AI/ML functions within the standard. Furthermore, legacy 802.11 mechanisms may need to be modified to integrate AI/ML native operations. It should be noted that the technical use cases defined by the AIML TIG are well aligned with the corresponding efforts of 3GPP, which in Rel-18 focuses on enhancements to data collection and signaling (to support AI/ML-based network energy savings, load balancing, and mobility optimization) and lists promising areas of interest for AI/ML implementations (e.g., channel state information, beam management, and positioning~\cite{lin2024overview}).

The following is an outline of the use cases defined by the AIML TIG ~\cite{aiml_tig_use_cases} (Figure~\ref{fig:use_cases}), including channel state information (CSI) feedback compression using neural networks (NNs), enhanced roaming assisted by AI/ML, deep reinforcement learning (DRL)-based channel access, and multi-access point coordination (MAPC) driven by AI/ML.

\subsection{CSI Feedback Compression}


Beamforming can significantly improve Wi-Fi capacity and reliability, though it comes at the cost of increased overhead due to the exchange of CSI reports.
These overheads can be substantially reduced using ML. In particular, deep learning-based auto-encoder architectures can potentially reduce the dimensionality of channel state matrices. Alternatively, since CSI may be similar for stations located close to each other, clustering methods such as k-means could be helpful. 
In this case additional upfront signaling is required, 
but this can be later offset by the reduced sounding overheads. Thus, the main cost of introducing AI/ML-based feedback compression remains in the computational complexity of the ML methods.

\subsection{Enhanced Roaming Assisted by AI/ML}

Roaming decisions (i.e., the handover of a client station between APs) can be improved through ML by learning station movement patterns and predicting received signal strength indicator (RSSI) levels to determine when to
initiate a scanning procedure. 
Besides the appropriate computational resources available at the AP for model training and execution, such a solution would require additional signaling in the form of extending existing 802.11 management frames (such as 802.11k neighbor report frames or 802.11v handover management frames).

\subsection{DRL-based Channel Access}

Numerous ML-based modifications to the 802.11 distributed channel access method have been proposed in the literature to improve its operation \cite{szott2022wifi}, but recently, solutions using reinforcement learning methods aim at learning optimal parameter values, e.g., contention window (CW) settings, for given network conditions. 
The report in~\cite{aiml_tig_use_cases} emphasizes that fair channel access conditions must be ensured for all stations, including legacy ones. Additionally, the computational complexity and the associated overhead should be minimized.

\subsection{MAPC Driven by AI/ML}

MAPC involves coordinating the transmissions of multiple APs in overlapping networks and is planned for the upcoming 802.11bn amendment~\cite{imputato2024beyond}. 
ML can support coordinated transmissions in multi-AP scenarios by either distributed or centralized selection of non-interfering AP-station pairs. 
ML-supported MAPC may require additional signaling, but this can be reduced if the implemented AI/ML model predicts network and traffic conditions.

\section{Towards a native, pervasive AI in IEEE 802.11 WLANs: The evolution path for Wi-Fi}
\label{sec:vision}

\begin{figure*}[t!]
    \centering    \includegraphics[width=0.9\textwidth]{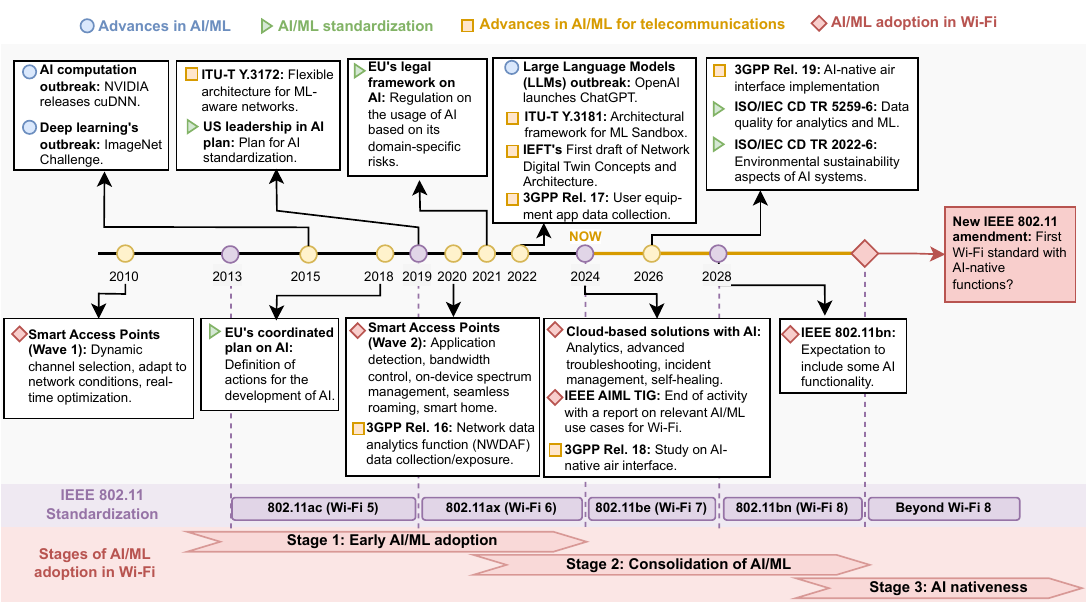}
    \caption{Roadmap of Wi-Fi towards AI nativeness.}
    \label{fig:roadmap}
\end{figure*}

The AIML TIG technical report described above does not finalize the standardization activities within the 802.11 since the AIML standing committee (SC) will continue reviewing and analyzing the feasibility and need for AI/ML in Wi-Fi. From now on, the AIML SC will serve as a liaison point on AI/ML topics, which can lead to new collaborations with other standards-developing organizations (SDOs) like ITU, ETSI, or 3GPP, who have already worked on AI/ML standardization for several years~\cite{shehzad2022artificial, ouyang2021next}, thus boosting innovation and bringing AI/ML adoption in Wi-Fi closer. Next, we overview the main challenges and enablers for the adoption of AI/ML in Wi-Fi and depict a possible roadmap towards AI nativeness. 

\subsection{AI/ML in Wi-Fi: Standardization Challenges}

\subsubsection{Backward compatibility}

One Wi-Fi-specific challenge is the backward-compatible premise of 802.11, whereby AI/ML features should take into account legacy devices, thus ensuring fairness in access to shared resources. Hence, AI/ML solutions must be properly designed and exhaustively tested to ensure that the AI/ML-operating devices are not disrupting legacy devices. For that, entities on top of 802.11 would be required to be involved, including the Wi-Fi Alliance, which is in charge of Wi-Fi certifications and which would need to introduce novel testing and certification track procedures for AI/ML. Alternatively, new greenfield bands can be used for AI/ML-based protocols, hence not disrupting legacy devices operating in different bands. This approach, however, entails involving spectrum regulators like ITU or FCC and requires strongly motivating the release of novel bands. 

\subsubsection{ML computations and overheads}

The data-driven condition of AI/ML methods makes them unquestionably computationally intensive, thus leading to extra computation and communication needs, as well as increased power consumption. In terms of computation, hardware-constrained devices would need to handle most of the ML operations, including data processing and ML model training and inference. Therefore, either their computational capabilities would need to be significantly extended, or computational support would be required (e.g., from the cloud). In terms of storage and communication, ML requires from tens to hundreds of megabytes. Moreover, depending on the training and inference approach (e.g., distributed learning), the overheads may dramatically increase due to the communication among ML components. Therefore, the use of ML models may turn out to be unfeasible if the computation and communication requirements are unbearable.

\subsubsection{Continuous development and integration}

ML model performance highly depends on the data with which the models are trained, thus requiring continuous monitoring and updating. Such volatility calls for flexible ML model deployment options, so that models can be updated at the targeted hardware on a plug-and-play basis. The plug-and-play approach has strong implications for the way 802.11 operates, as it would require the unification and opening of rooted interfaces on the chipsets. With standardized ML-specialized procedure interfaces, vendors could implement their own AI/ML-based functionalities on top. This is aligned with 3GPP's vision for Rel-18, where the definition of AI/ML models is expected to be part of the implementation and left out of the standard~\cite{RP221348}. For example, for data rate selection, the 802.11 only defines transmission rates available, thus the actual selection can be done by proprietary ML methods.

\subsubsection{Interoperability}

The heterogeneity of ML solutions can cause conflicts and lead to fairness issues, especially if ML optimizations involve channel access. Additionally, if ML is applied to manipulate how information is transmitted (e.g., build fields in transmitted frames), interoperability issues may arise between devices from different vendors. In this regard, it is important from the point of view of 802.11 standardization to clearly define the bounds of operation for AI/ML and to standardize new fields, so that the underlying principles of Wi-Fi (e.g., fair channel access) are guaranteed and interoperability is provided. 
Furthermore, the developed AI/ML methods will need to adhere to spectrum regulations, including existing rules, such as ETSI's standards on channel access in European unlicensed bands, and upcoming spectrum sharing in the newly opened 6~GHz bands: both vertical (with incumbent users) and horizontal (planned sharing with other services). Finally, AI/ML methods also have the potential to impact future methods of spectrum management, an area in which there is a need for novel solutions on a global scale, provided that such methods are trustworthy.
In this regard, the Federal Communications Commission (FCC) recently called for new ideas and means for performing spectrum management, including those based on AI/ML \cite{fcc2023}.

\subsubsection{General concerns on AI/ML}

Besides networking-specific challenges, AI/ML may raise broad concerns in the areas of security, privacy, explainability, and ethics~\cite{leslie2019understanding}. AI has progressed at a fast pace in recent years, even faster than regulations on AI use. As a downside, AI may cause harm due to a large set of reasons (e.g., lack of explainability, misuse, poor data used for training, bad design), thus leading to misbehavior. As a result, AI may lead to unexpected insecure actions, thus potentially compromising security for mission-critical applications. Other major threats of AI concern cybersecurity and privacy. The fact that AI relies on large data and often requires networking (e.g., to support ML pipelines distributed among different locations or collaborative ways of model training) entails the need for significant data transfers, often involving sensitive information. Moreover, the proliferation of specialized parties with expertise in AI/ML may lead to marketplaces for trading raw ML models, trained models, or training data, which can further contribute to cybersecurity flaws.

\subsection{Keys for Successful Adoption of AI/ML in Wi-Fi}

To fulfill both the TIG use cases presented above and future use cases, 802.11 will need to extend its operation to accommodate the required ML functionalities. We next identify key aspects for the successful adoption of AI/ML in Wi-Fi.

\subsubsection{Feasibility analysis}

For the adoption of particular AI/ML use cases, it is required to exhaustively validate their performance and behavior and to demonstrate their superiority against currently adopted mechanisms (e.g., autoencoder-based CSI compression vs codebook-based compression). First, relevant KPIs (e.g., throughput, packet error rate) must be identified and studied using standard-compliant tools such as simulators or testbeds. In this regard, tools integrating 802.11 simulations with AI/ML functions, e.g., ns-3-gym~\cite{gawlowicz2018ns3}, will gain a lot of importance. The validation of AI/ML solutions and their generalization capabilities also depends on the data used and on the trust placed in particular ML models (often acting as black boxes). In this regard, the standardization of datasets and evaluation scenarios (cf. the work in progress ISO/IEC CD TR 5259-6) can be of great utility, as well as the integration of explainable AI tools and digital twins (cf. ITU-T Y.3181) in the validation process. Apart from the performance analysis, AI/ML solutions must be assessed from an implementation point of view. For that, it is of utmost importance to understand their needs in terms of data requirements and their potential impacts on the specification (e.g., the additional signaling required by AI/ML to fulfill a particular task).

\subsubsection{Implementation considerations}

AI/ML processes---data collection, data processing, model deployment, model training/inference, ML output application---need to be considered to extend the current 802.11 protocols and devices~\cite{wilhelmi2020flexible}. As a first step towards AI/ML implementations, relevant SDOs have already provided guidelines and elements to support AI/ML operations related to data collection and processing (cf. the 3GPP  network data analytics function and the ITU-T Y.3174 specification). In Wi-Fi, AI/ML processes such as data collection or model sharing will also require architectural changes, so that such functionalities can be enabled, for instance, on the MAC layer of APs and stations. The need for exchanging ML models may lead to the proliferation of ML marketplaces (cf. ITU-T Y.3176). Regarding training and inference operations, new signaling would be required to indicate support for such capabilities (e.g., in federated learning, ML model weights need to be exchanged during the training of a model). Finally, the computation complexity and time to run AI/ML processes (e.g., inference), together with the use case requirements, will determine the hardware and software needs of the executing device. For that, additional means for indicating computation capabilities might be required within 802.11, so that ML operations can be unlocked. When it comes to cybersecurity and privacy, 802.11 must provide the necessary authentication and encryption means to protect the data and information related to AI/ML operation, so that attacks such as data/model poisoning or eavesdropping can be avoided. On top of that, data storage must meet the necessary security and privacy requirements (e.g., through encryption and anonymization), which may require efforts outside the 802.11 standard.

\subsection{AI/ML in Wi-Fi: Adoption and Standardization Roadmap}

Knowing that AI/ML can improve Wi-Fi's performance and unlock novel use cases and functionalities, it is crucial to examine how the IEEE 802.11 standard can sustain such evolution. Owing to the challenges listed above, the journey towards AI-native Wi-Fi is not straightforward, and will only be realized as a result of several years of research and joint efforts in a complex regulatory, standardization, and industrial ecosystem. In the AI/ML in communications roadmap, we identify three potential stages of adoption (defined in the sequel and highlighted on the bottom of Fig.~\ref{fig:roadmap}), based on the degree of integration and involvement of AI/ML in the Wi-Fi protocol.

\subsubsection{Early AI/ML adoption} 

At this stage, AI/ML solutions developed on top of the 802.11 standard provide support for operating and moderately enhancing Wi-Fi networks. One type of early adopted solution consists of cloud-based ML applications whereby Internet service providers (typically in the order of millions of subscribed customer premise equipment) leverage a vast amount of data from their Wi-Fi subscribers to improve the way their network is operated (e.g., enhanced network management and troubleshooting). Cloud-based solutions rely on the data collected by the providers through, for instance, USP-based solutions, which allow retrieving AP and station measurements (e.g., based on 802.11k neighbor reports). The data is then processed in the cloud using powerful and specialized equipment in data centers (e.g., graphics/tensor processing units). The main challenges of cloud-based AI/ML solutions lie in the collection of massive data, which is often constrained (e.g., allowing updates every 10 or 15 minutes), and in the time complexity of running the AI/ML models. At the early AI/ML adoption stage, we also find low-complexity solutions such as those running in limited smart APs/stations (e.g., intelligent channel selection, radio resource management, and link adaptation),  which are thoroughly described in~\cite{szott2022wifi}. These solutions typically entail minimum challenges in terms of computation and storage due to their simplicity. 


\subsubsection{Consolidation of AI/ML} 

The latest efforts in 802.11 standardization (cf. Section \ref{sec:standardization}) and the proliferation of Wi-Fi products with AI/ML (e.g., Qualcomm's AI-optimized Wi-Fi 7 chipset FastConnect 7900) underscore the willingness to integrate AI/ML deeper into Wi-Fi. Such efforts suggest that the 802.11 specification can be updated in the following years to allow the integration of vendor-specific AI/ML features (e.g., ML-based CSI compression, channel access, and coordination). 
At this stage, computationally constrained devices (APs and stations) will be required to perform AI/ML-based operations, especially regarding data collection and full or partial inference. As a result, the main challenges of this phase stem from the space complexity. For that reason, any vendor-based AI/ML solutions implemented on the Wi-Fi chipsets are required to leverage already available data as much as possible (e.g., legacy CSI reports). Additionally, special care should be taken regarding efficient memory utilization and selection of AI/ML algorithms.

\subsubsection{AI nativeness} 

The last stage of adoption envisages AI/ML built-in 802.11 PHY and MAC features. This stage goes beyond the current way of developing standards and solutions (building self-AI/ML protocols~\cite{rial2023role}), so its success depends on many advances in the regulation, improvement, trustworthiness, and optimization of AI/ML. From the standardization point of view, AI nativeness can be achieved by expanding the current protocols to support AI/ML-specific operations, thus leading to new procedures and interfaces to enable ML pipelines (cf. 3GPP's AI-native air interface~\cite{RP221348}). To fulfill AI nativeness, Wi-Fi devices will likely require additional computation and storage capabilities to give response to novel functionalities (e.g., self-learned MAC protocols). Moreover, advances towards efficient AI (e.g., neuromorphic computing, ML optimization, tiny ML) are foreseen to be key enablers for AI nativeness in Wi-Fi.

\section{Use Case Results: From On-Top Intelligence to AI-Native WLANs}
\label{sec:reults}

\begin{figure}[ht!] 
    \centering
    \includegraphics[width=\linewidth]{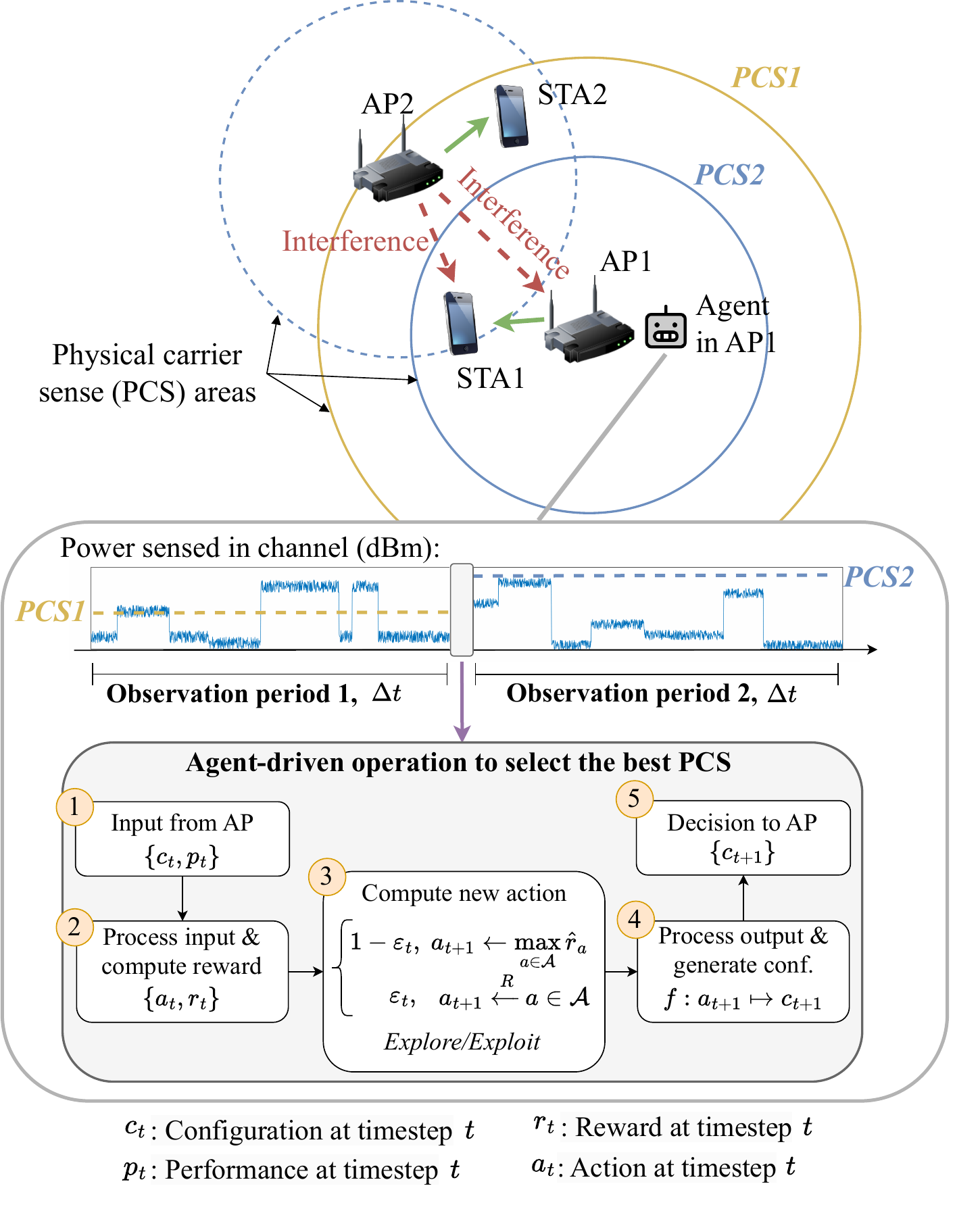} 
    \caption{Agent-based operation in 802.11 networks.}
    \label{fig:use_case_description}
\end{figure}

To showcase the benefits that AI/ML could bring to Wi-Fi at different stages of adoption, we study the use case of ML-based spatial reuse (SR). Here, the AI/ML operation is embodied by agents implementing multi-armed bandits algorithms (e.g., $\varepsilon$-greedy), which  derive the best SR configuration to be used by each basic service set (BSS). In particular, an agent adheres to each AP, thus forming a multi-player setting (i.e., with competing multiple agents) in deployments with multiple APs~\cite{wilhelmi2019potential}. The agents' operation is depicted in Fig.~\ref{fig:use_case_description}, where agents retrieve configuration and performance inputs from the APs (step \#1), process them to compute a reward (step \#2), derive a new action based on a given policy (step \#3), process the action to generate a new configuration (step \#4), and communicate the decision to the AP (step \#5). To capture different degrees of maturity for AI/ML adoption, we consider:
\begin{enumerate}
    \item \textbf{Constrained  SR} (\texttt{11axSR}): The agent is constrained to the SR operation defined in IEEE 802.11ax, so it learns the best carrier sensing threshold, $C\in[-82, -62]$. Note that the selection of $C$ entails a transmit power limitation, where the maximum allowed transmit power is inversely proportional to $C$.
    \item \textbf{Unconstrained SR} (\texttt{Free}): The agent freely adjusts both the carrier sensing and the transmit power to maximize performance. Through this approach, the fairness in channel access is implicit in the agents' rewards, thus in their expected behavior~\cite{wilhelmi2019potential}. Unconstrained SR is backward-compatible with legacy devices as long as current spectrum regulations are respected (e.g., a clear channel assessment, CCA, check is performed). Nevertheless, higher performance could potentially be achieved if regulations are revised to rely on the self-adaptation capabilities of AI/ML.
\end{enumerate}


Moreover, with the advent of MAPC in Wi-Fi~8, we define two types of rewards to be used by the agents: decentralized (\texttt{DEC}), whereby each agent attempts to maximize its throughput (no coordination), and coordinated (\texttt{COORD}), whereby agents share their performance to maximize the max-min throughput. The proposed use case is simulated using Komondor, an 802.11 simulator which includes agent operation, with the parameters given in Table~\ref{tab:sim_parameters}.\footnote{The source files used in this paper are available at \url{https://github.com/mlwifitutorial/towards_ai-native_wifi}. Accessed on July 16, 2024.}

\begin{table}[ht!]
\centering
\caption{Simulation parameters.}
\label{tab:sim_parameters}
\resizebox{\columnwidth}{!}{%
\begin{tabular}{@{}lll@{}}
\toprule
 & \textbf{Parameter} & \textbf{Value} \\ \midrule
\multirow{5}{*}{\rotatebox[origin=c]{90}{\textit{Scenario}}} & Num. of random deployments, $N$ & $100$ \\
 & Deployment side size, $D$ & $20$~m \\
 & Number of BSSs, $N_\text{BSS}$ & $4$ \\
 & Traffic model, $\lambda$ & Full-buffer \\
 & Simulation time, $T$ & $100$~s \\
 \midrule
\multirow{8}{*}{\rotatebox[origin=c]{90}{\textit{PHY/MAC}}} &  Central frequency, $f_c$ & $6$~GHz \\
 & Bandwidth, $B$ & $20$~MHz \\
 & Default transmit power, $P_\text{default}$ & $20$~dBm \\
 & Default carrier sensing, $S_\text{default}$ & $-82$~dBm \\ 
 & Min. contention window, $CW_\text{min}$ & $16$ \\
 & Max. contention window stage, $s_\text{max}$ & $5$ \\
 & Maximum aggregated MPDUs, $N_{agg}$ & $64$ \\
 & Data frame length, $L$ & $12$~kbit \\
 \midrule
\multirow{5}{*}{\rotatebox[origin=c]{90}{\textit{Agents}}} & Transmit power levels, $\mathbf{P}$ & $\{5,10,15,20\}$~dBm \\
 & Carrier sensing levels, $\mathbf{C}$ & $\{-82,-78,-74,-70,-66,-62\}$~dBm \\
 & Action-selection strategy & $\varepsilon$-greedy \\ 
 & Initial $\varepsilon$, $\varepsilon_0$ & 1 \\ 
 & Exploration adaptation over time $t$, $f_\varepsilon(\cdot)$ & $\varepsilon_t = \varepsilon_0 / \sqrt{t}$ \\
 \bottomrule
\end{tabular}%
}
\end{table}

\begin{figure}[ht!] 
    \centering
    \includegraphics[width=\columnwidth]{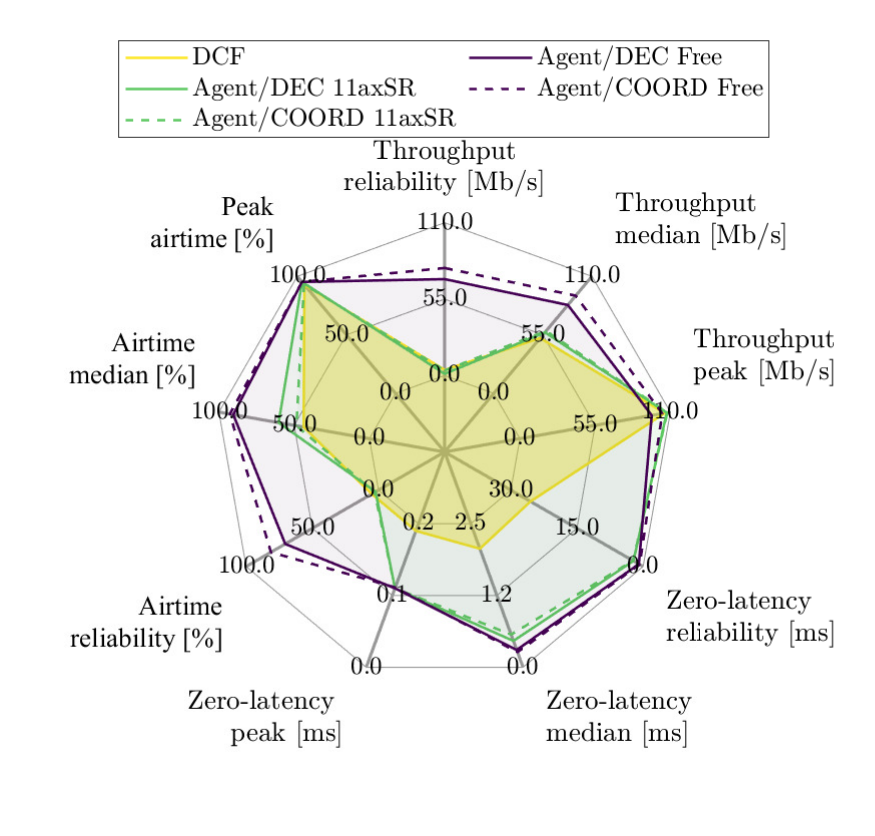}
    \caption{Performance achieved by each approach.}
    \label{fig:spiderplot}
\end{figure}

The results achieved by each approach are shown in Fig.~\ref{fig:spiderplot}. They are compared against the baseline without agents (\texttt{DCF}) for three relevant metrics, namely throughput, latency, and airtime. In all presented cases, reliability (25th percentile), median (50th percentile), and peak (75th percentile) are provided. As shown, the \texttt{11axSR}-constrained agent approach provides only slightly better results than legacy \texttt{DCF} in terms of throughput, which can be attributed to the intrinsic limitations of the 802.11ax SR specification, which is conservative by design. Notwithstanding, \texttt{11axSR} agents allow to significantly enhance the zero-latency reliability by selecting the proper value of $C$. When it comes to the unconstrained agents approach (\texttt{Free}), we observe that the reliability and median gains substantially outperform both \texttt{DCF} and \texttt{11axSR} settings, thus proving the potential of an AI-native approach. 

Regarding agent coordination through MAPC, \texttt{COORD} provides slightly more benefits in the \texttt{Free} case, where agents have more freedom in accessing the channel and adapting the power to be used, than in the \texttt{11axSR} case, where any configuration $C$ intrinsically includes fairness thanks to the protection mechanisms within 802.11ax SR. In the former case (\texttt{Free}), \texttt{COORD} improves \texttt{DEC} by 10\%, 11\%, and 8\% in terms of reliability, median, and peak throughput, respectively. When it comes to \texttt{11axSR}, \texttt{COORD} and \texttt{DEC} lead to very similar performances (e.g., 58.63~Mb/s vs 57.30~Mb/s in terms of  median throughput or 2.28~ms vs 2.19~ms in terms of zero-latency reliability), thus \texttt{DEC} is the most feasible option in this case due to its lower implementation cost (i.e., communication between agents is not required).



Considering that the room for improvement for SR is somewhat limited in dense scenarios like the one showcased in this paper due to high interference regimes, the performance achieved by the unconstrained AI/ML-driven SR turns out to be promising and serves as an illustrative example for motivating AI-native communications. Moreover, the presented experiments are based on decentralized multi-armed bandits, thus the gains provided by AI/ML could potentially be higher if more sophisticated solutions were considered. 
While the coordinated setting (\texttt{COORD}) with local decision-making leads to only minor improvements compared to the non-coordinated setting (\texttt{DEC}), it enables a more gentle context whereby different APs can collaborate and enforce network-wise policies (e.g., proportional fairness).



\section{Conclusions}
\label{sec:conclusions}

We have shown the ongoing standardization efforts and outlined a tentative roadmap towards future AI/ML-native Wi-Fi interfaces. Based on the provided list of related challenges, we conclude that the success and speed of development of AI/ML-native Wi-Fi interfaces will mainly depend on future standardization efforts to ensure the effectiveness, feasibility, and trustworthiness of the solutions developed by Wi-Fi vendors. This standardization process will require great care, as it is necessary to not only define appropriate interfaces precisely but also possible constraints (e.g., sets of parameters/functions/modules that can be modified) and security mechanisms. Overall, the definition and development of AI/ML-native Wi-Fi is still in its early stages and will certainly require answers to many open questions, however, it promises tempting performance improvements.

\ifCLASSOPTIONcaptionsoff
\newpage
\fi

\bibliographystyle{IEEEtran}
\bibliography{bib}

\end{document}